\documentclass{article}

\usepackage{PRIMEarxiv}

\usepackage[utf8]{inputenc} 
\usepackage[T1]{fontenc}    
\usepackage{hyperref}       
\usepackage{url}            
\usepackage{booktabs}       
\usepackage{amsfonts}       
\usepackage{nicefrac}       
\usepackage{microtype}      
\usepackage[linesnumbered,ruled,vlined]{algorithm2e}
\usepackage{fancyhdr}       
\usepackage{graphicx}       
\usepackage{amsmath}
\usepackage{amsfonts}
\usepackage{balance}
\usepackage{braket}
\usepackage{xcolor}

\title{Cutting Quantum Circuits to Run on Quantum and Classical Platforms}

\author{
  Wei Tang\\
  Department of Computer Science \\
  Princeton University \\
  \texttt{weit@princeton.edu}
   \And
  Margaret Martonosi \\
  Department of Computer Science \\
  Princeton University \\
  \texttt{mrm@princeton.edu}
}

\begin{document}
\maketitle

\begin{abstract}
  Quantum computing (QC) offers a new computing paradigm that has the potential to provide significant speedups over classical computing.
  Each additional qubit doubles the size of the computational state space available to a quantum algorithm.
  Such exponentially expanding reach underlies QC's power,
  but at the same time puts demanding requirements on the quantum processing units (QPU) hardware.
  On the other hand, purely classical simulations of quantum circuits on either central processing unit (CPU) or graphics processing unit (GPU) scale poorly
  as they quickly become bottlenecked by runtime and memory.
  This paper introduces CutQC,
  a scalable hybrid computing approach that distributes a large quantum circuit onto quantum (QPU) and classical platforms (CPU or GPU) for co-processing.
  CutQC demonstrates evaluation of quantum circuits that are larger than the limit of QPU or classical simulation,
  and achieves much higher quantum circuit evaluation fidelity than the large NISQ devices achieve in real-system runs.
\end{abstract}


\section{INTRODUCTION}\label{sec:introduction}
QC has emerged as a promising computational approach with
the potential to benefit numerous scientific fields.
For example, some of the earliest QC work shows that
quantum algorithms for factoring~\cite{shor1999polynomial} can be exponentially faster than their classical counterparts.
However, these quantum algorithms assume the existence
of large-scale, fault-tolerant, universal quantum computers.

Instead, today's quantum computers are noisy intermediate-scale quantum (NISQ) devices.
Major challenges limit their effectiveness.
Noise can come from limited coherence time,
frequency selection for individual qubits,
crosstalk among qubits,
and limited control bandwidth.
Because of these and other issues,
the difficulty of building reliable quantum computers increases dramatically with increasing number of qubits.

More fundamentally, such intermediate-scale quantum devices are hard limited by their qubit count.
Currently, only small quantum circuits can be run on small quantum computers.
The largest superconducting quantum computers available today have $127$ qubits,
and their relatively poor fidelity further limits the size of circuits that can be reliably run.

Both the noise and the intermediate-scale characteristics of NISQ devices present significant obstacles to their practical applications.
On the other hand, the alternative for quantum circuits evaluation---classical simulations of quantum circuits---produces noiseless output but is not tractable in general.
For example, state-of-the-art classical simulations of quantum circuits of $100$ qubits require $42$ million cores~\cite{liu2021closing}.

This work uses circuit cutting to
expand the reach of small quantum computers with partitioning and post-processing techniques 
that augment small QPU platforms with CPUs and GPUs.
CutQC is an end-to-end hybrid approach that automatically locates efficient cut positions to cut a
large quantum circuit into smaller subcircuits that are each independently executed by QPUs with less quality and size requirements.
Via scalable post-processing techniques,
the output of the original circuit can then be reconstructed or sampled efficiently from the subcircuit outputs with classical computing.

To evaluate the performance of CutQC,
we benchmarked four different quantum circuits that represent a general set of circuits for gate-based QC platforms and promising near-term applications.
We demonstrate executing quantum circuits of up to 100 qubits on existing NISQ devices and classical computing.
This is significantly beyond the current reach of either quantum or classical methods alone.

Our contributions include the following:
\begin{enumerate}
    \item \textbf{Expanding the size} of quantum circuits that can be run on NISQ devices and classical simulation by combining the two.
    Our method allows executions of quantum circuits more than twice the size of the available quantum computer backend and much beyond the classical simulation limit.
    \item \textbf{Improving the fidelity} of quantum circuit executions on NISQ devices.
    We show an average of $21\%$ to $47\%$ improvement to $\chi^2$ loss for different benchmarks by using CutQC with small QPUs, as compared with direct executions on large QPUs.
\end{enumerate}
\section{BACKGROUND}\label{sec:background}
\begin{figure}[t]
    \centering
    \includegraphics[width=1\linewidth]{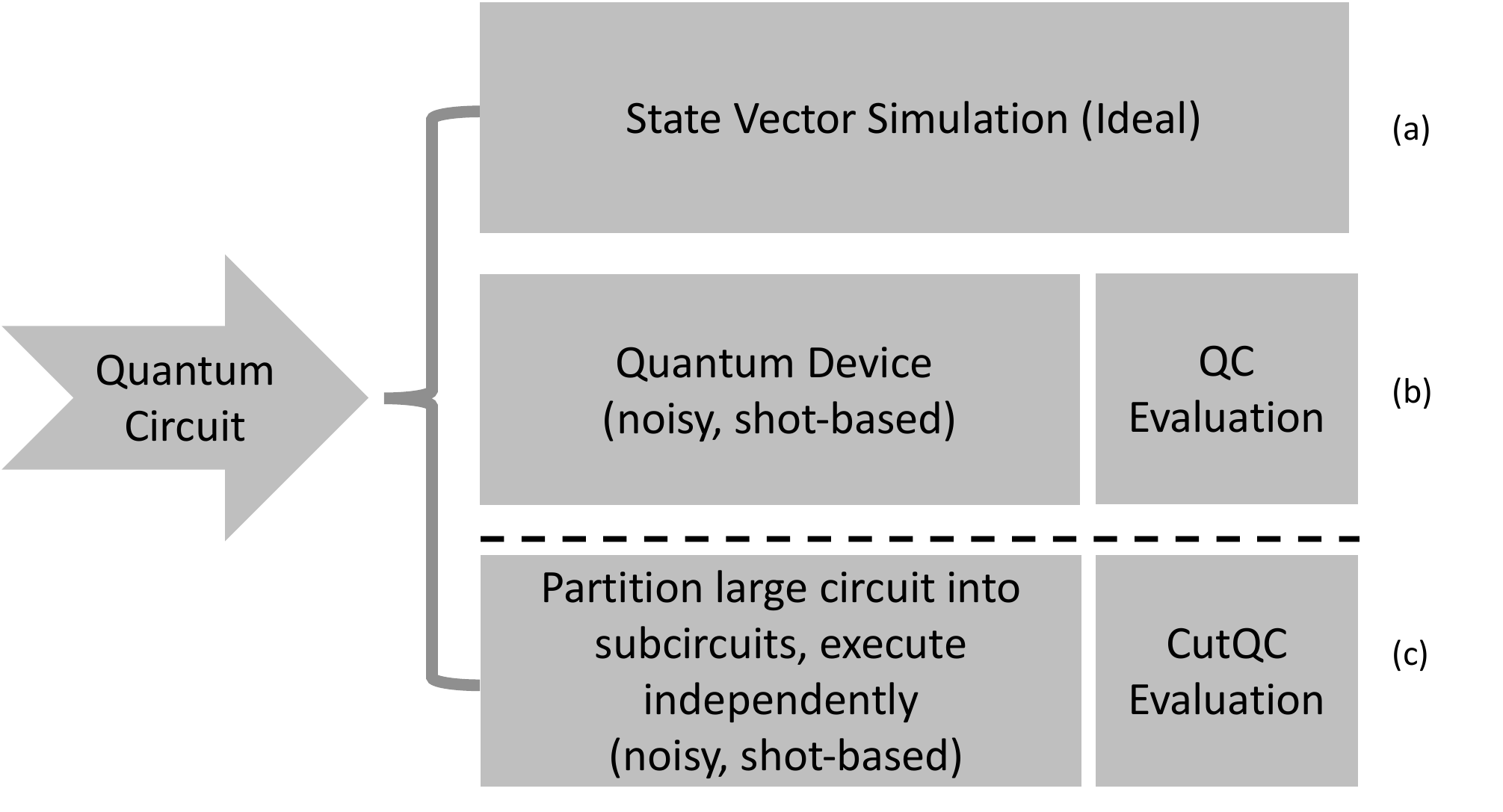}
    \caption{Different quantum circuit evaluation modes.
    (a) Purely classical simulation produces the ground truth to verify other evaluation outputs.
    (b) Purely quantum evaluation on a single QPU.
    Multiple vendors provide cloud access to their devices.
    (c) Our hybrid mode, which is orders of magnitude faster than (a),
    produces much less noisy outputs than (b),
    and evaluates much larger circuits than (a) and (b).}
    \label{fig:eval_modes}
\end{figure}
This section introduces quantum circuits and explains the differences between several quantum circuit evaluation modes.

Quantum programs are expressed as circuits that consist of a sequence of single- and multiqubit gate operations.
Quantum circuits can be evaluated by using classical simulations, on quantum computers, or in a hybrid mode as explored in this paper.
Figure~\ref{fig:eval_modes} provides an overview of the different evaluation modes.

State vector simulation (Figure~\ref{fig:eval_modes}a) is an idealized noiseless simulation of a quantum circuit.
All quantum operations are represented as unitary matrices.
N-qubit operations are $2^N\times2^N$ unitary matrices.
State vector simulation executes circuits by sequentially multiplying each gate's corresponding unitary matrix with the current state vector.
This yields an error-free output represented as complex amplitudes,
which cannot be obtained on quantum computers.
This evaluation mode scales exponentially and serves to provide the ground truth and runtime baselines for benchmarking NISQ devices for small quantum circuits.
We use this evaluation mode as a baseline to verify the output of modes (b) and (c) in Figure~\ref{fig:eval_modes} and to compute the $\chi^2$ metric to quantify the noise and quality of quantum circuit executions.

QC evaluation (Figure~\ref{fig:eval_modes}b) physically executes quantum circuits on NISQ computers using a shot-based model.
Quantum algorithms are first compiled to satisfy device-specific characteristics such as qubit connectivity, native gate set, noise, and crosstalk.
A real NISQ device then executes the compiled quantum circuit thousands of times (``shots'') in quick succession.
At the end of each shot, all qubits are measured; and the output, a classical bit string, is recorded.
After all shots are taken, a distribution of probabilities over the observed states is obtained.

This paper explores the CutQC evaluation (Figure~\ref{fig:eval_modes}c) that combines both quantum and classical platforms.
Section~\ref{sec:results} demonstrates the runtimes of the CutQC evaluation (Figure~\ref{fig:eval_modes}c) for large quantum circuits on smaller QPUs.
We also compare the execution fidelities of the QC evaluation (Figure~\ref{fig:eval_modes}b) and the hybrid evaluation (Figure~\ref{fig:eval_modes}c) modes.
\section{CIRCUIT CUTTING}
While we refer the readers to~\cite{peng2020simulating} for a proof of the physics theory behind cutting quantum circuits,
this section provides an intuitive understanding of the cutting process and its challenges via an illustrative example.

\subsection{Circuit Cutting: Example}\label{sec:cc_example}
\begin{figure*}[t]
    \centering
    \includegraphics[width=.95\textwidth]{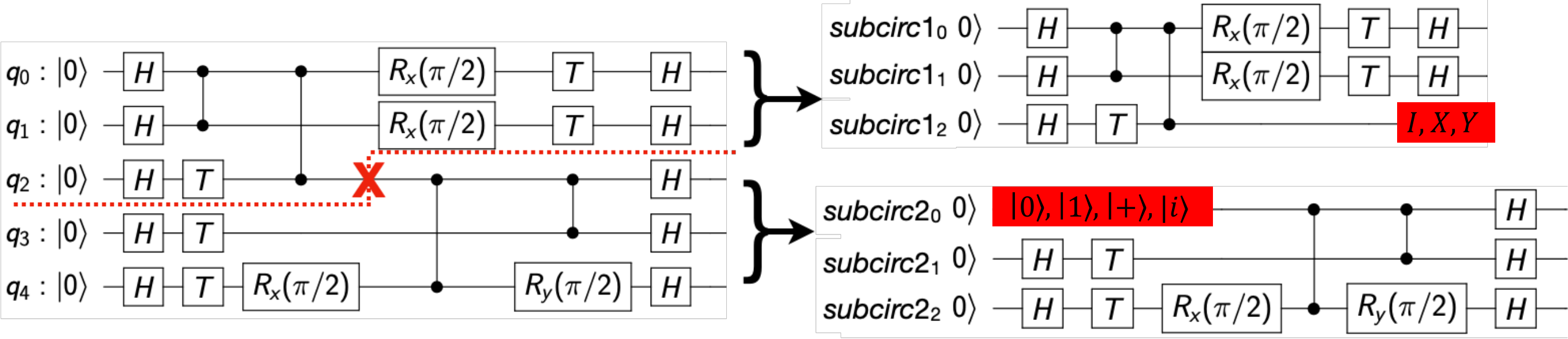}
    \caption{Example of cutting a $5$-qubit circuit into two smaller subcircuits
    of $3$ qubits each. The subcircuits are produced by cutting the
    $q_2$ wire between the first two $cZ$ gates.
    The three variations of $subcircuit_1$ and four variations of $subcircuit_2$ can then be evaluated on a 3-qubit QPU, instead of a 5-qubit QPU.
    The classical postprocessing involves summing over $4$ Kronecker products between the two subcircuits for the one cut made.}
    \label{fig:cutting_example}
\end{figure*}

Consider the quantum circuit example in Figure~\ref{fig:cutting_example}.
One cut separates a 5-qubit quantum circuit into 2 subcircuits of 3 qubits each.
Time goes from left to right in quantum circuit diagrams,
and each row represents a qubit wire.
CutQC performs vertical cuts on qubit wires, in other words, timewise cuts.
The qubit states across the cutting point are then decomposed into their Pauli bases.

With a proper selection of the cutting points, a large quantum circuit can be divided into smaller isolated subcircuits.
Without cutting, the circuit in Figure~\ref{fig:cutting_example} at least requires a $5$-qubit QPU with good enough qubits to execute all the quantum gates before too many errors accumulate.
Circuit cutting divides this quantum circuit and produces two smaller subcircuits, each with both fewer qubits and fewer gates.
Now multiple less powerful 3-qubit QPUs can run these independent subcircuits in parallel.
The quantum interactions among the subcircuits are substituted by classical post-processing,
which are analogues to the communication cost paid in classical parallel computing.

In general, a $n$ qubit quantum circuit undergoes $K$ cuts to divide into $n_C$ completely separated subcircuits $C = \left\{C_1,\ldots,C_{n_C}\right\}$.
A complete reconstruction of the quantum interactions requires each cut to permute each of the Pauli $\{I,X,Y,Z\}$ bases, for a total of $4^K$ combinations.
Depending on the Pauli basis assigned to each cut, the subcircuits are initialized and measured slightly differently to produce a distinct entry.
We use $p_{i,k}$ to represent the output of subcircuit $i$ in the $k$th edge bases assignment, where $i\in\{1,\ldots,n_C\}$ and $k\in\{1,\ldots,4^K\}$.
The physics theory dictates that the output of the original circuit is given by
\begin{equation}
    P=\sum_{k=1}^{4^K}\otimes_{i=1}^{n_C}p_{i,k}\in\mathbb{R}^{2^n}\label{eq:CutQC}
\end{equation}
where $\otimes$ is the tensor product between two subcircuit output vectors.
\subsection{Circuit Cutting: Challenges}\label{sec:cc_challenges}
The first challenge is to find cut locations.
While quantum circuits can always be split into smaller ones,
finding the optimal cut locations is crucial in order to minimize the classical postprocessing overhead.
In general, large quantum circuits may require more than one cuts in order to be separated into subcircuits.
In this case, the cutting scheme evaluates all possible measurement-initialization combinations.
The resulting number of Kronecker products is $4^K$, where $K$ is the number of edges cut.
For general quantum circuits with $n$ quantum edges,
this task faces an $\mathcal{O}(2^n)$ combinatorial search space.
Section~\ref{sec:MIP} addresses this problem with mixed-integer programming.
Our work shows that with only a few cuts, many useful applications can be tractably mapped to NISQ devices currently available.

The second challenge is to scale the classical postprocessing.
Large quantum circuits have exponentially increasing state space that quickly becomes intractable to even store the full-state probabilities.
Section~\ref{sec:DD_post_processing} addresses this problem with a dynamic definition algorithm to efficiently locate the ``solution'' states
or sample the full output distribution for large quantum circuits beyond the current QC and classical simulation limit.
\section{Framework Overview}\label{sec:framework}
\setlength{\belowcaptionskip}{-5pt}
\begin{figure}[t]
    \centering
    \includegraphics[width=.6\linewidth]{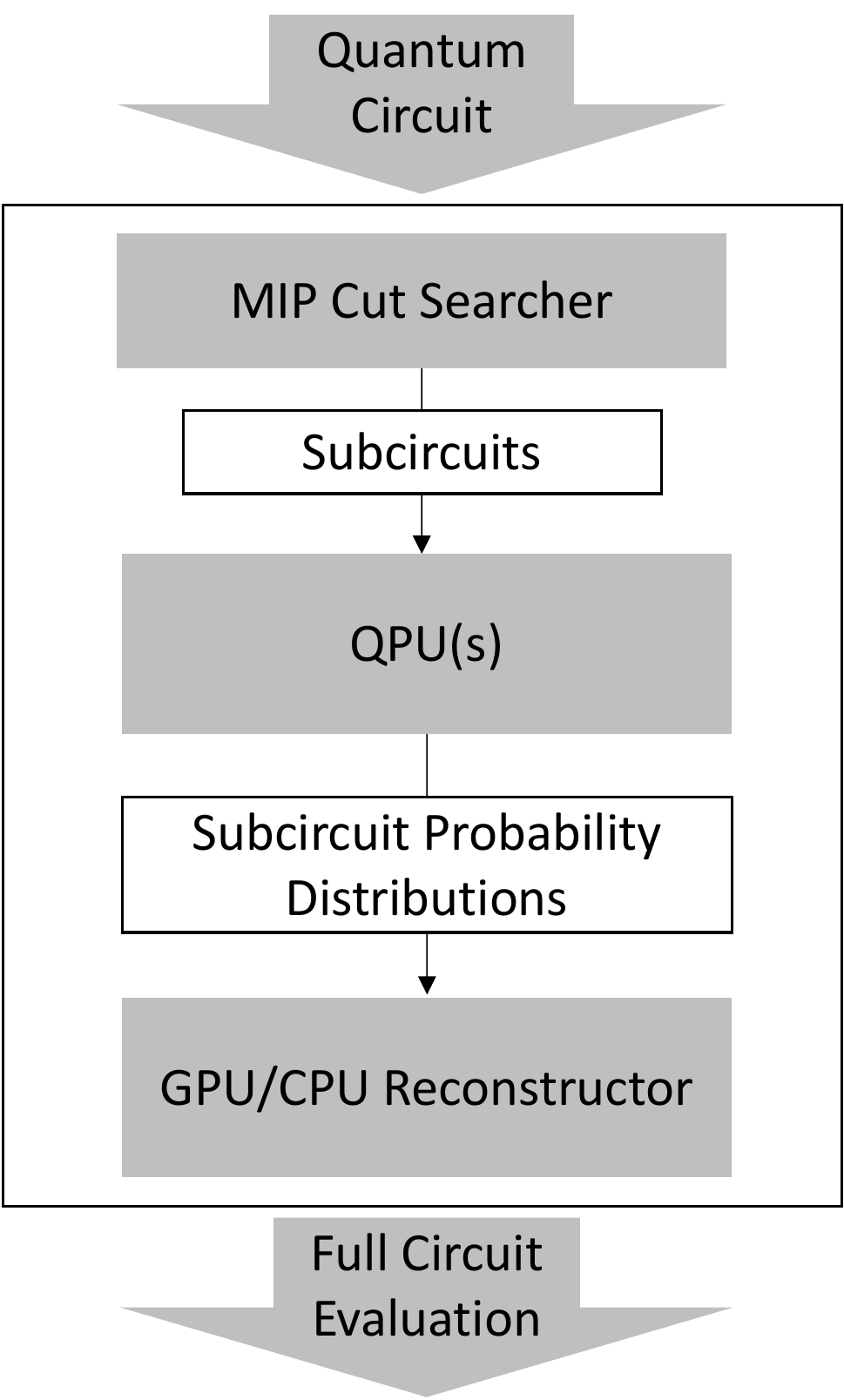}
    \caption{Framework overview of CutQC. A mixed-integer programming (MIP) cut searcher
    automatically finds optimal cuts given an input quantum circuit.
    The small subcircuits resulting from the cuts are then evaluated by using quantum devices.  The reconstructor then reproduces the probability distributions of the original circuit.}
    \label{fig:cutting_mode}
\end{figure}
Figure~\ref{fig:cutting_mode} summarizes the key components of our framework.
CutQC is built on top of IBM's Qiskit package in order to use IBM's quantum devices for the experiments on fidelity,
but we note that the hybrid approach works with any gate-based quantum computing platforms.
Given a quantum circuit specified as an input,
the first step is to decide where to make cuts.
We propose the first automatic scheme that uses mixed-integer programming to find optimal cuts for arbitrary quantum circuits.
The backend for the MIP cut searcher is implemented in the Gurobi solver.
Multiple QPUs then evaluate the different combinations of the subcircuits.
Eventually, a reconstructor running on either CPUs or GPUs postprocesses the subcircuit outputs
and reproduces the original full circuit outputs from the Kronecker products.
\subsection{MIP Cut Searcher}\label{sec:MIP}
Unlike the manual example in Section~\ref{sec:cc_example},
CutQC's cut searcher uses mixed-integer programming (MIP) to automate the identification of cuts that require the least amount of classical postprocessing.
Our problem instances are solved by the Gurobi mathematical optimization solver~\cite{gurobi}.

The framework assumes that the input quantum circuit is fully connected.
That is, all qubits are connected via multiqubit gates either directly or indirectly through intermediate qubits.
A quantum circuit that is not fully connected can be readily separated into fully connected subcircuits without cuts,
and does not need the classical postprocessing techniques to sew together.
We hence focus on the more difficult general cases where cutting and reconstruction are needed.

We adopt the public MIP solver from~\cite{tang2021cutqc},
which solved the constrained partition problem by predicting the post-processing to directly compute Equation~\ref{eq:CutQC}.
Besides the input quantum circuit,
the MIP cut searcher also requires the user to specify
(1) the maximum number of qubits allowed per subcircuit,
and (2) the maximum number of subcircuits allowed.
(1) is just the size of the quantum devices available to the user.
(2) is set to $5$ in this paper.

Locating the cut points is equivalent to clustering the multi-qubit gates in the input quantum circuit.
A quantum circuit can be modeled as a directed acyclic graph (DAG).
Quantum operations are always applied sequentially to the qubits.
The single-qubit gates are ignored during the cut-finding process,
since they do not affect the connectivity of the quantum circuit.
The multi-qubit quantum gates are then modeled as the vertices,
and the qubit wires are modeled as the edges.
Choosing which edges to cut in order to split the circuit into subcircuits is equivalent to clustering the vertices.
The corresponding cuts required to produce the clustering are hence the cross-cluster edges.

We seek to minimize the classical postprocessing overhead required to reconstruct a circuit from its subcircuits.
Therefore, the objective is set to be the number of floating-point multiplications involved in computing Equation~\ref{eq:CutQC},
given by:
\begin{equation}\label{eq:MIP_obj}
    L\equiv4^K\sum_{c=2}^{n_C}\prod_{i=1}^c2^{n_i}.
\end{equation}
where $K$ is the number of cross-cluster edges, i.e. the number of cuts.
$n_C$ is the number of subcircuits,
and $n_i$ is the number of qubits in subcircuit $i$.
This cost objective accurately captures the bulk of the computation when we aim
to build the full $2^n$ probabilities for an $n$-qubit uncut circuit, under the
full definition CutQC mode (discussed in Section~\ref{sec:FD_post_processing}). 

However, there is a prohibitive memory requirement for storing the $2^n$
probabilities as floating-point numbers when circuits get larger.
Section~\ref{sec:DD_post_processing} introduces a novel dynamic definition
method to efficiently sample very large circuits with a much lower
postprocessing overhead. Nevertheless, we chose to minimize
Equation~\ref{eq:MIP_obj} during cut search as a positively correlated
objective.
\subsection{Full Definition Post-Processing}\label{sec:FD_post_processing}
We developed two types of classical postprocessing algorithms: a full-definition (FD) query and a dynamic-definition (DD) query algorithms.
The difference in these methods lies in whether the entire $2^n$ full-state probability output of the uncut circuit is reconstructed.

The reconstruction step (computing Equation~\ref{eq:CutQC}) is essentially taking vector-vector tensor products.
Previous work~\cite{tang2021cutqc} used Intel CPUs as the classical backends,
and demonstrated significant runtime advantages of hybrid computation over classical simulations in the full state setting.
Since GPUs are particularly suitable for inter vector tensor products,
this paper runs the classical post-processing on a single GPU via Tensorflow.
\subsection{Dynamic Definition Post-Processing}\label{sec:DD_post_processing}
Quantum circuits can be loosely categorized into two groups.
The first group produces sparse output probabilities,
where just a few ``solution'' states have very high probabilities
and the ``non-solution'' states have low or zero probabilities.
Most known quantum algorithms fall into this category,
such as Bernstein--Vazirani algorithm~\cite{bernstein1997quantum} and the Quantum Fourier Transform (QFT)~\cite{cooley1965algorithm}.
This is where QC shows promise over classical computing by efficiently locating the ``solution'' states.

The second group of circuits produces dense output probabilities,
where many states have nonzero probabilities.
For this type of circuit,
even with access to QPUs large enough to execute the circuits directly,
querying the FD probability output quickly becomes impossible.
The reasons are that
(1) an exponentially increasing amount of memory is required to store the probabilities
and (2) an exponentially increasing number of shots are required on a QPU before the probabilities converge.
Fortunately, knowing the FD probabilities of all states simultaneously is usually not of interest.
Instead, users are interested in the distribution itself.

DD query allows us to find the ``solution'' states
or sample dense probability distributions efficiently with very large quantum circuits,
even when storing the full-state probability is not tractable.
DD query produces a probability distribution that merges certain states into one bin and maintains the sum of their probabilities instead of the individual states within.

Algorithm~\ref{alg:dynamic_definition} presents the DD algorithm.
In each recursion, DD runs the subcircuits to produce the merged subcircuit outputs before post-processing.
The \textit{active} qubits in each recursion determine the number of bins,
the \textit{merged} qubits determine which states are merged into the same bin,
and the \textit{zoomed} qubits indicate the qubit states that have been fixed.
Each subsequent recursion zooms into the bin with the largest sum of probability from the previous recursions,
improving the `definition' of the states contained in the bin.
This lets DD recursively obtain more fine-grained outputs for the input circuit.

\begin{algorithm}[t]
    \DontPrintSemicolon
    \SetAlgoVlined
    \caption{Dynamic Definition}\label{alg:dynamic_definition}
    \KwIn{Subcircuits from cutting.\;
    Max number of qubits that fit in the memory per recursion $M$.\;
    Max number of recursions $R$.}
    Initialize an empty list $L$\;
    $r\gets0$\;
    \While{$r<R$}{
        \uIf{$r=0$}{
            Choose a maximum of $M$ qubits to label as ${\color{red}active}$\;
        }
        \uElse{
            Fix the quantum states of the ${\color{red}active}$ qubits in $bin$\;
            Label the qubits as ${\color{blue}zoomed}$\;
            }
        Label the rest of the qubits as ${\color{orange}merged}$\;
        \textbf{QPUs} : Run the subcircuits to produce the sum of probabilities for the subcircuit bins
        by grouping shots with the same ${\color{red}active}$ qubits quantum states together\;
        Reconstruct the $2^{\#{\color{red}active}}$ probability output for the ${\color{red}active}$ qubits\;
        Append the $R$ largest bins still with ${\color{orange}merged}$ qubits to $L$\;
        Sort and truncate $L$ to keep the largest $R$ bins\;
        Pop $bin$ from $L$\;
        $r\gets r+1$\;
        }
\end{algorithm}

For sparse outputs, DD recursively pinpoints the ``solution'' states and their probabilities.
To do so, DD query follows a DFS-like search strategy to recursively choose the $bin$ with higher probabilities to zoom in on.
By recursively locating the $active$ qubits in their most probable $zoomed$ states,
``solution'' states can be easily located after just a few recursions.
For an $n$-qubit full circuit, the number of recursions needed is $\mathcal{O}(n)$.

For dense outputs, DD builds a ``blurred'' probability landscape of the exact FD probability distribution,
with the ability to arbitrarily ``zoom in'' on any region of the state space.
To do so, DD query follows a BFS-like strategy to choose the $bin$ with higher probabilities to zoom in on.
This is equivalent to efficient sampling of very large circuits on less powerful QPUs and less memory.

\section{Methodology}\label{sec:methodology}
This section introduces the various backends, metrics and benchmarks for the experiments.
\subsection{Backends}
We test our approach by running post-processing and classical simulation benchmarks on both CPUs and GPUs.
The CPU backend comprises of Intel(R) Xeon(R) Platinum 8260 CPUs at 2.40GHz, with $256$ GB allocated memory.
We tested on two single-node CPU settings, one with $16$ CPUs and another with $64$ CPUs.
The GPU backend is a single Nvidia A100 GPU.

\subsection{Metrics}
The CutQC runtime is the end-to-end runtime except the QPU time in Algorithm~\ref{alg:dynamic_definition}.
This is because the NISQ QPUs nowadays are small, slow and too noisy for any practical purposes.
The applications of CutQC to useful algorithms at large scales requires medium sized reliable QPUs instead.
It is hence irrelevant to profile the NISQ QPU runtime now.
Furthermore, we expect that the QPU runtime will be negligible as compared to the other parts of the toolflow because
(1) QPUs operate at much faster timescales than post-processing on CPUs and GPUs,
and (2) multiple small QPUs can be used in parallel to reduce the runtime.
In addition, the runtime advantage of QPUs over CPUs will be even more significant for larger circuits.
We expect CutQC to offer more significant advantages over purely classical methods as larger and more reliable QPUs become available.

In addition, we profile the output fidelity of CutQC with IBM's 5-qubit Bogota device to compare the fidelity with directly executing the circuits on IBM's 20-qubit Johannesburg device.
As NISQ devices improve, CutQC can be applied to larger devices to produce useful executions on larger scales.
To quantify the noise behaviors, we used $\chi^2$ loss 
\begin{equation}
    \chi^2=\sum_{i=0}^{2^n-1}\frac{(a_i-b_i)^2}{a_i+b_i}\label{eq:chi2},
\end{equation}
where $a_i$ are elements of circuit execution probability distributions (from Figure~\ref{fig:eval_modes}b,~\ref{fig:eval_modes}c)
and $b_i$ are elements of the ground truth probability distributions (from Figure~\ref{fig:eval_modes}a).
The smaller the $\chi^2$ is, the better the execution results are.

\subsection{Benchmarks}
We used the following circuits as benchmarks.
\begin{enumerate}
    \item \textit{Bernstein--Vazirani} (\textit{BV}).
    This quantum algorithm solves the hidden string problem more efficiently than classical algorithms do~\cite{bernstein1997quantum}.
    \item \textit{Adder}.
    Adder is a quantum ripple-carry adder with one ancilla and linear depth.
    It is an important subroutine in quantum arithmetic involving summing two quantum registers of the same width;
    hence only even numbers of qubits are valid.
    \item \textit{Approximate Quantum Fourier Transform} (\textit{AQFT}).
    QFT is a common subroutine in many quantum algorithms that promise speedup over classical algorithms.
    AQFT has been proposed to yield better results than QFT on NISQ devices by truncating small angle rotations~\cite{barenco1996approximate}.
    \item \textit{Supremacy}.
    This is a type of 2-D random circuit with dense probability output.
    It was used by Google to demonstrate quantum advantage~\cite{google2019quantum}.
    The circuit depth is $10$ in our experiments.
    We verified that the rectangular shapes (such as $2*10$) are much easier to be cut and require little postprocessing.
    We therefore focused only on the more difficult near-square shapes, with the two dimensions differing by up to $2$ qubits (such as $4*5$).
    Hence not all numbers of qubits are valid.
\end{enumerate}
The benchmark circuits represent a general set of circuits for gate-based QC platforms and promising near-term applications.

\subsection{Summary of Experiments}
Previous work has demonstrated significant runtime advantages of the CPU implementations over classical simulations in the FD settings~\cite{tang2021cutqc},
we hence focus on comparing the performance of GPUs versus CPUs in the DD settings for large circuits.
We tested DD query for circuits up to $100$ qubits,
significantly beyond the current classical and quantum limit.
Because no backends are capable of producing accurate circuit executions on this scale,
we used random numbers as the subcircuit output to focus on studying the runtime.

In addition, we tested running circuits in the FD mode on a $5$-qubit IBM QPU,
and compared the output fidelity against direct QC evaluations on a $20$-qubit IBM QPU.
\section{EXPERIMENT RESULTS}\label{sec:results}
\subsection{Dynamic Definition Query}\label{sec:dynamic_definition_results}
\setlength{\belowcaptionskip}{-5pt}
\begin{figure}[t]
    \centering
    \includegraphics[width=\linewidth]{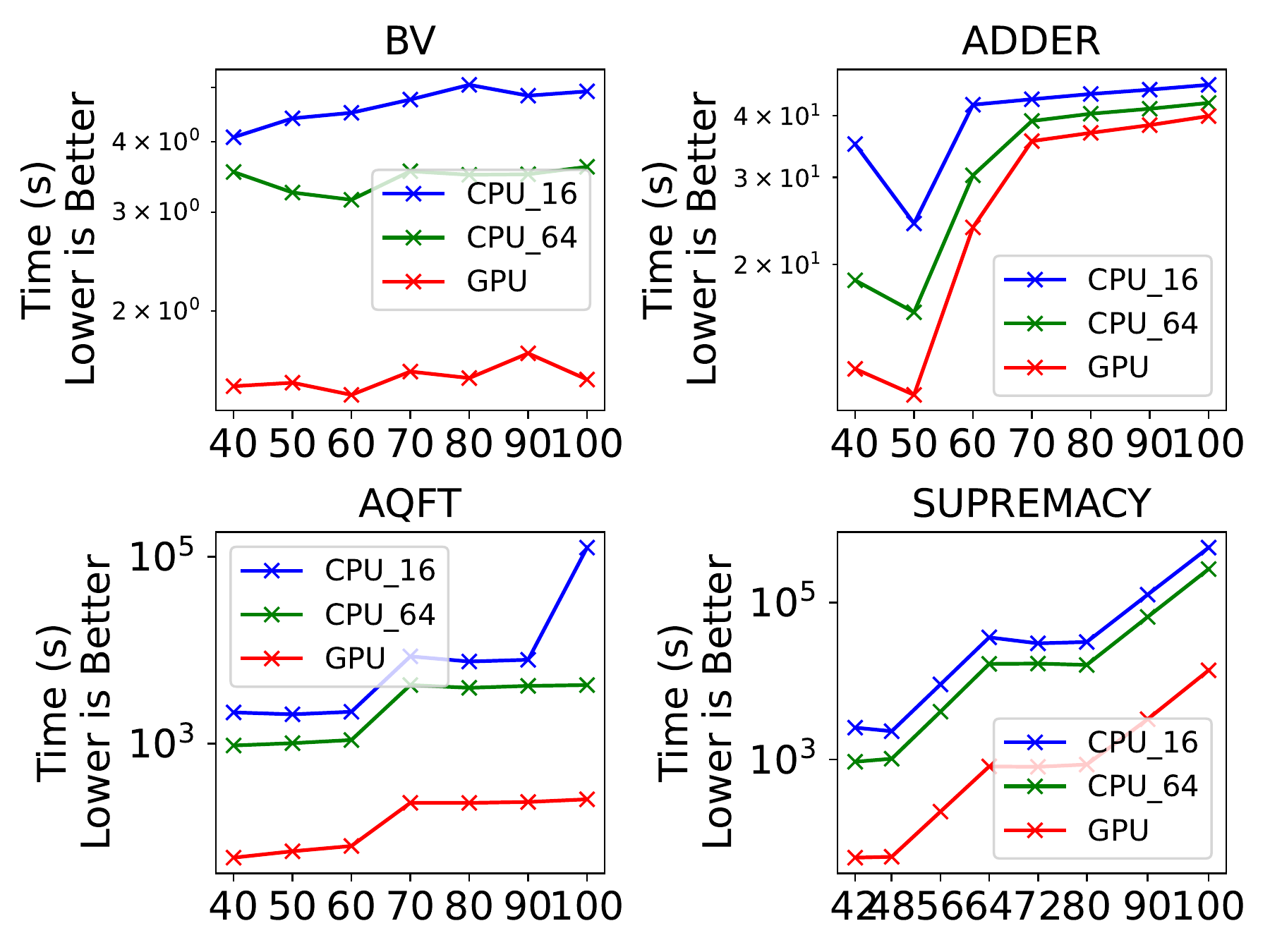}
    \caption{CutQC executes benchmark circuits mapped to quantum devices with up to $\frac{3}{4}$ of the number of qubits in the input circuits.
    The horizontal axis shows the number of qubits in the input circuit.
    The vertical axis shows the postprocessing runtime of $1$ DD recursion with a definition of $2^{30}$ bins.
    GPU is the fastest backend as expected.}
    \label{fig:dd_expand}
\end{figure}

We used DD to efficiently sample quantum circuits of which the full Hilbert space is too large to even store.
NISQ devices will gradually improve in fidelity and sizes to allow evaluating subcircuits beyond the classical simulation limit.
CutQC will then allow the use of those NISQ devices to efficiently evaluate even larger quantum circuits.
We cut and executed circuits of up to $100$ qubits and used DD query to sample their blurred probability landscape with a definition of $2^{30}$ bins in one recursion.

Figure~\ref{fig:dd_expand} shows the runtime of cutting and mapping circuits to quantum computers with up to $\frac{3}{4}$ of the qubits.
The classical post-processing overhead in FIgure~\ref{fig:dd_expand} is hence the classical `cost' to expand the reach of QPUs by at least a quarter more of the qubits available.
Certain benchmarks, such as BV, almost double the number of qubits possible via CutQC.
Furthermore, the novel incorporation of GPUs makes such cost minimal
to gain the huge benefit of significantly expanding the reach of the underlying quantum and classical platforms alone.
In fact, GPU provides up to two orders of magnitude runtime improvements in benchmarks that are harder to cut and hence require more classical post-processing,
such as \textit{AQFT} and \textit{Supremacy}.
This is all without the need for either a large quantum computer or vast classical computing resources.

Note that neither the CPU or the GPU backends used in the experiments alone is capable of running any of the benchmark circuits in Figure~\ref{fig:dd_expand}.
\subsection{Real QC Runs}\label{sec:device_results}
\setlength{\belowcaptionskip}{-5pt}
\begin{figure}[t]
    \centering
    \includegraphics[width=.9\linewidth]{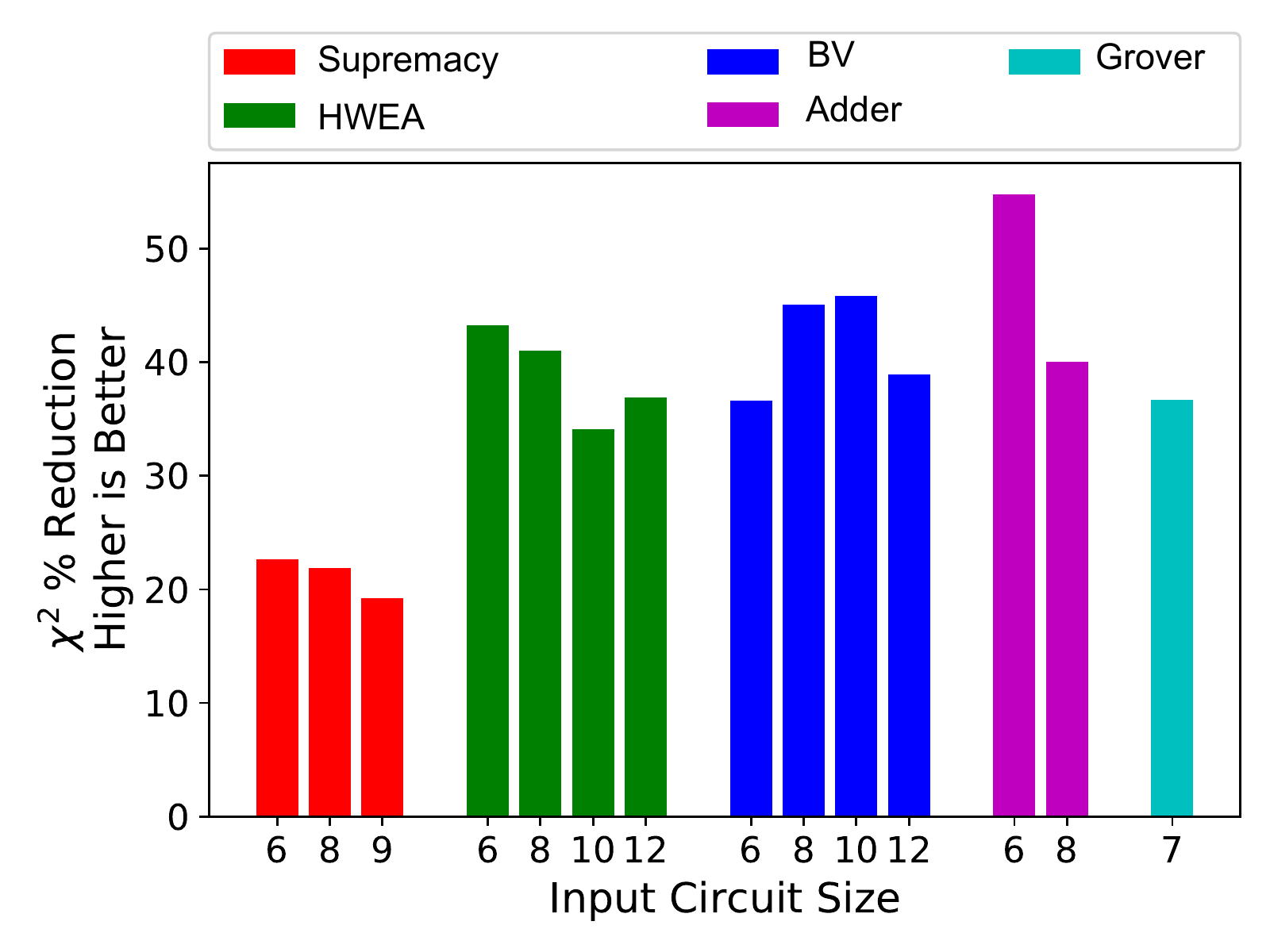}
    \caption{Comparison of the 20-qubit Johannesburg quantum computer versus the 5-qubit Bogota device with CutQC.
    For each benchmark we find the ideal output distribution via statevector simulation.
    We then use this ideal distribution to compute the $\chi^2$ metric for two execution modes:
    QC evaluation on the Johannesburg device ($\chi^2_J$) and CutQC evaluation utilizing the Bogota device ($\chi^2_B$).
    The reported $\chi^2$ percentage reduction is computed as $100*(\chi^2_J - \chi^2_B) / \chi^2_{J}$.
    A distribution that is close to ideal will have a small $\chi^2$ value,
    and therefore a positive $\chi^2$ percentage reduction indicates improved performance.
    Only the AQFT workloads experience a negative reduction and are omitted.
    CutQC achieves an average of $21\%$ to $47\%$ $\chi^2$ reduction for different benchmarks.}
    \label{fig:chi2}
\end{figure}

To study the effect of device noise on our toolchain,
we ran experiments on IBM's real quantum devices.
Figure~\ref{fig:chi2} compares the circuit output obtained from
(a) directly executing circuits on the state-of-the-art 20-qubit Johannesburg device
and (b) executing circuits with more than 5 qubits on the 5-qubit Bogota device with CutQC.
We show that CutQC evaluation with small quantum computers produces a lower $\chi^2$ loss and hence outperforms QC evaluation with large quantum computers.
CutQC reduces $\chi^2$ loss by nearly $60\%$ in the best cases.
The experiments stop at $12$ qubits because QC evaluation beyond this point succumbs to the effects of noise and fails to produce meaningful output.
Among the benchmarks, only the AQFT circuits experienced a negative reduction.
This is because AQFT compiled for the current NISQ devices is much deeper than the other benchmarks.
Therefore both QC and CutQC on AQFT have accuracy too low for meaningful comparisons.
As NISQ devices improve in noise and connectivity, we expect AQFT to improve.

Despite requiring more subcircuits and readout,
CutQC evaluates circuits with better fidelity.
The main reason for such improvements is that CutQC runs subcircuits that are both smaller and shallower than the uncut circuit run by the QC mode.
Furthermore, CutQC substitutes the noisy quantum entanglement across subcircuits by noise-free classical postprocessing.

Not only does CutQC need smaller quantum computers,
it also produces better outputs.
Therefore, combined with CutQC, building small but reliable quantum computers becomes much more useful than merely increasing qubit counts at the cost of degrading fidelity.
\section{Related Work}\label{sec:related_works}
Many quantum compilation techniques have been developed to improve the performance of NISQ devices.
However, these focus on improving a purely quantum computing approach and are intrinsically limited by the size and fidelity of NISQ devices.
Specifically, our experiments used the noise adaptive compiler~\cite{murali2019noise} in both CutQC and QC evaluations.
The improved fidelity we demonstrate is in addition to that given by the compiler.
Furthermore, previous compilers do not allow executions of circuits beyond quantum computer sizes at all.
Our approach can work in concert with any compilers to execute circuits both larger in size and better in fidelity.

Previous works on classical simulation require massive computing resources,
or only simulate very few output states at a time~\cite{liu2021closing}.
Many small-scale quantum circuit cutting demonstrations exist for chemical molecule simulations~\cite{eddins2021doubling}
and variational quantum solvers~\cite{yuan2021quantum}.
\section{Conclusion}\label{sec:conclusion}

This paper demonstrates how to leverage both quantum and classical computing platforms together to execute quantum algorithms of up to $100$ qubits while simultaneously improving the fidelity of the output.
Our results are significantly beyond the reach of current quantum or classical methods alone, and our work pioneers pathways for scalable quantum computing.
Even as NISQ machines scale to larger sizes and as fault-tolerant QPUs emerge,
CutQC's techniques for automatically cutting and efficiently reconstructing quantum circuit executions offer a practical strategy for hybrid quantum/classical advantage in QC applications.
\section*{Code Availability}
Our codes are available at: https://github.com/weiT1993/CutQC.

\section*{Acknowledgements}
This work is partly funded by EPiQC, an NSF Expedition in Computing, under grants CCF-1730082/1730449.
This work is partly based upon work supported by the U.S. Department of Energy, Office of Science, National Quantum Information Science Research Centers, Co-design Center for Quantum Advantage (C2QA) under contract number DE-SC0012704.

\bibliographystyle{unsrt}  
\bibliography{references}

\end{document}